\documentclass{ws-ijmpd}
\usepackage[super,compress]{cite}

\usepackage{amsmath}
\usepackage{graphicx,floatrow,array}
\usepackage{textcomp}

\newcommand{\unit}[1]{\:\mathrm{\textstyle #1}}

\newcommand{\Lag}{\mathcal{L}}
\newcommand{\K}{\mathcal{K}}

\let\blqq=\textquotedblleft
\let\brqq=\textquotedblright

\makeatletter
\newcommand{\drm}{\@ifstar{\mathrm{d}}{\:\mathrm{d}}}
\makeatother

\graphicspath{{Bilder/}}

\begin{document}

\markboth{Arne Grenzebach, Volker Perlick, Claus L\"ammerzahl}
{Photon Regions and Shadows of Accelerated Black Holes}

%
\catchline{}{}{}{}{}
%

\title{PHOTON REGIONS AND SHADOWS OF ACCELERATED BLACK HOLES}

%
%

\author{ARNE GRENZEBACH\footnote{arne.grenzebach@zarm.uni-bremen.de},
VOLKER PERLICK\footnote{volker.perlick@zarm.uni-bremen.de}, and
CLAUS L\"AMMERZAHL\footnote{claus.laemmerzahl@zarm.uni-bremen.de}}
\address{ZARM, University of Bremen, \\
Am Fallturm, 28359 Bremen, Germany}

\maketitle

\begin{history}
\received{Day Month Year}
\revised{Day Month Year}
\end{history}

\begin{abstract}
In an earlier paper we have analytically determined the photon regions 
and the shadows of black holes of the Pleba{\'n}ski class of metrics
which are also known as the Kerr--Newman--NUT--(anti-)deSitter metrics.
These metrics are characterized by six 
parameters: mass, spin, electric and magnetic charge, 
gravitomagnetic NUT charge, and the cosmological constant. Here
we extend this analysis to the Pleba{\'n}ski--Demia{\'n}ski class
of metrics which contains, in addition to these six parameters,
the so-called acceleration parameter. All these metrics are 
axially symmetric and stationary type D solutions to the Einstein--Maxwell 
equations with a cosmological constant.  
We derive analytical formulas for the photon regions (i.e., for the
regions that contain spherical lightlike geodesics) and for the 
boundary curve of the shadow as it is seen by an observer at 
Boyer--Lindquist coordinates $(r_O, \vartheta _O)$ in the domain of 
outer communication. Whereas all relevant formulas are derived
for the whole Pleba{\'n}ski--Demia{\'n}ski class, we concentrate 
on the accelerated Kerr metric (i.e., only mass, spin and acceleration
parameter are different from zero) when discussing the influence of
the acceleration parameter on the photon region and on the shadow
in terms of pictures. The accelerated Kerr metric is also known 
as the rotating $C$-metric. We discuss how our analytical formulas
can be used for calculating the horizontal and vertical angular 
diameters of the shadow and we estimate these values for the black
holes at the center of our Galaxy and at the center of M87.
\end{abstract}


\ccode{PACS numbers: 04.70.-s, 95.30.Sf, 98.35.Jk}
%
%
%
%


\section{Introduction}
\label{sec:intro}
Basically, the \emph{shadow} of a black hole is the region on the 
observer's sky that is left dark if the light sources are anywhere
in the universe but not between the observer and the black hole. 
For a mathematical description, it is convenient to consider light
rays that are sent from the observer's position into the past.
Some of them are deflected by the black hole and then go out to bigger
radii again; because they can reach one of the light sources, we assign 
brightness to their initial directions. Others stay 
close to the radial line and go to the horizon without meeting one of 
the light sources; to their initial directions we assign darkness. 
The resulting dark region on the sky is called  the
\emph{shadow} of the black hole. The boundary of the shadow is 
determined by light rays that spiral towards a lightlike geodesic 
which stays on a sphere. The region outside of the black hole filled 
with these spherical lightlike geodesics is called the exterior 
\emph{photon region}. Recently, the shadow of a black has attracted 
even Hollywood's attention. In the movie \emph{Interstellar}, 
which was released in 2014, it was shown how nearby observers would 
see the shadow of an almost extremal rotating (Kerr) black hole with 
an accretion disk\cite{JamesTunzelmann.2014}.

In the near future astronomers actually expect to observe the shadow of a black hole.
Currently there are two cooperating projects---the US-led Event Horizon 
Telescope project\cite{DoelemanWeintroub.2008} and the European 
BlackHoleCam project---who try to image the shadow of galactic black holes.
It has been predicted since several years that there are supermassive 
black holes at the centers of most---if not all---galaxies.  There
is strong evidence for a black hole at the center of our own Galaxy,
associated with the radio source Sagittarius~A* (Sgr~A*):
Infrared observations of orbits of stars near the 
center\cite{EckartGenzel.1996,GillessenEisenhauerEtAl.2009} demonstrate
that there has to be a heavy object with a mass of approximately
$4.3$ million Solar masses within a small volume. The most convincing 
candidate for such an object is a black hole. Another good candidate for 
a supermassive black hole is the object at the center of M87 
($m>10^{9}$ Solar masses). Because of the large distance, the 
diameter of the shadow of galactic black holes will be tiny. Even for the 
nearest candidate Sgr~A* 
($8.3\unit{kpc}$ away\cite{GhezSalim.2008,GillessenEisenhauerEtAl.2009}),
the predicted diameter of the shadow is only about
$50\unit{microarcseconds}$, see Sec.~\ref{sec:angrad}. Although tiny, such 
a diameter should be resolvable with very large baseline interferometry 
(VLBI)\cite{DoelemanWeintroub.2008,HuangCai.2007}. Numerical studies
taking scattering into account suggest that the shadow can be 
observed only at sub-millimeter wavelengths, see Falcke, Melia and 
Agol\cite{FalckeMelia.2000}. 
The resolution of interferometric measurements in this wavelength regime 
will be further improved if the 10-meter space-based radio-telescope 
\emph{Millimetron}\cite{KardashevNovikov.2014} goes 
into operation, probably in the mid-2020s.
With this Russian satellite the Earth-based telescope network is upgraded 
with an extra-long Space-Earth baseline of $1.5\unit{million\, km}$. 

If the shadow of a black hole will be observed, its shape will give 
important information on the parameters of the black hole. Therefore, it 
is necessary to provide the observers with theoretical calculations of the 
shape of the shadow for different black-hole models. In an earlier
paper~\cite{GrenzebachPerlick.2014}, we have given an analytic
formula for the shadow of black holes of the Pleba{\'n}ski class, which are
also known as Kerr--Newman--NUT--(anti-)de Sitter space-times. Black holes
of this class are characterized by mass, spin, electric or 
magnetic charge, a NUT-parameter and a cosmological constant. In the
present paper we extend this analysis to the bigger class of 
Pleba{\'n}ski--Demia{\'n}ski space-times, which include in addition
a so-called acceleration parameter. The Pleba{\'n}ski--Demia{\'n}ski 
space-times are axially symmetric stationary solutions to the 
Einstein--Maxwell equations with a cosmological constant and they 
are of Petrov type D. Before determining the boundary
curve of the shadow in these space-times, we have to study the photon 
region. We develop the relevant formulas for the whole class of 
Pleba{\'n}ski--Demia{\'n}ski space-times. However, when studying 
the effect of the acceleration parameter onto the photon region and 
onto the shadow in terms of pictures, we restrict to black holes for 
which only the mass, the spin and the acceleration parameter are 
different from zero. This is because the effect of the other parameters 
-- electric and magnetic charge, NUT-parameter and cosmological constant 
-- has been studied already in our earlier paper and the presence of 
the acceleration does not change this significantly. The metric of these
black hole space-times characterized by mass, spin and acceleration alone 
is known as the rotating $C$-metric or the accelerated Kerr metric.

If only the mass and the acceleration parameter are different from 
zero, we have the so-called $C$-metric which describes a space-time
with boost-rotation symmetry. This solution to the vacuum Einstein
field equation was found by Levi-Civita 
(1919)\cite{Levi-Civita.1919} and Weyl (1919)\cite{Weyl.1917,Weyl.1919}.
The name $C$-metric refers to the classification in the
review of Ehlers and Kundt (1962)\cite{EhlersKundt.1962}. The rotating 
version of the $C$-metric was considered by Hong and Teo\cite{HongTeo.2005} 
while a detailed discussion of accelerated space-times in general can be 
found in the book by Griffiths and Podolsk{\'y}\cite{GriffithsPodolsky.2009}. 

Commonly the $C$-metric is given in the form introduced by Hong and 
Teo\cite{HongTeo.2003}
\begin{equation}
	g_{\mu \nu}^C \drm x^{\mu} \drm x^{\nu} = \frac{1}{\alpha^{2}(x+y)^{2}} 
	\Bigl( -F \drm \tau^2 + \frac{\drm y^{2}}{F} + 
		\frac{\drm x^{2}}{G} + G \drm \varphi^2 \Bigr)
\end{equation}
with cubic functions $G=(1-x^2)(1+2\alpha mx)$ and $F=-(1-y^2)(1-2\alpha my)$.
The metric depends on two 
parameters, the mass $m$ and the acceleration parameter $\alpha$.
The domain covered by the coordinates $(\tau, x,y,\varphi)$ 
actually contains \emph{two} black holes accelerating away from each 
other with a conical singularity (a ``strut'') on the axis of
rotational symmetry\cite{GriffithsPodolsky.2009,KinnersleyWalker.1970,
Bonnor.1983,BonnorDavidson.1992}. For our purposes, Boyer--Lindquist 
coordinates are more suitable, see 
Eq.~\eqref{eq:Metric} below, which cover only one of the two black holes.

There are several earlier papers on the shadows of black holes. Here we 
just mention some important works, for a more comprehensive list we refer 
to the introduction of our earlier paper\cite{GrenzebachPerlick.2014}. 
The first analytic calculations of the shadow of a black hole were done 
by Synge\cite{Synge.1966} for the Schwarzschild metric (Synge used
the word ``escape cone'' for the complement of the shadow) and by
Bardeen\cite{Bardeen.1973} for the Kerr metric. In the Schwarzschild 
case the photon region reduces to the ``photon sphere'' at $r=3m$
and the shadow is circular. In the Kerr case, the photon region is 
spatially three-dimensional and the shadow is non-circular. 
The deviation of the shadow from a circle could be used as a measure 
for the spin\cite{HiokiMaeda.2009}. %
With ray-tracing algorithms it is possible to include effects of matter
on the light rays like emission regions and 
scattering\cite{BardeenCunningham.1973,Luminet.1979,DexterAgol.2012,YounsiWu.2012,
MoscibrodzkaFalcke.2014,MoscibrodzkaShiokawa.2012,DexterFragile.2013}.
Here we do not take such effects into account but restrict ourselves
to the purely geometric construction of the shadow based on the assumption
that light rays are lightlike geodesics and that there are no light sources 
between us and the black hole. It is our goal to derive an analytical
formula for the shape of the shadow from which, in future work, the 
parameters of the black hole could be extracted with analytical means.

After a discussion of relevant properties of the 
Pleba{\'n}ski--Demia{\'n}ski space-times (Sec.~\ref{sec:PD}) we 
determine the photon region 
(Sec.~\ref{sec:regionK}) which is essential for calculating the  
boundary of the shadow of the black hole for an observer
at Boyer--Lindquist coordinates $(r_O, \vartheta_O)$ 
(Sec.~\ref{sec:shadow}). We derive all relevant formulas for the whole 
Pleba{\'n}ski--Demia{\'n}ski class. However, when illustrating the results
with pictures of the photon region and of the shadow in 
Secs.~\ref{sec:regionK} and \ref{sec:shadow} we restrict to the
accelerated Kerr metric.
In Sec.~\ref{sec:angrad} we discuss how our analytical formulas
can be used for calculating the angular diameters of the shadow. We 
use these equations for estimating the angular diameters of the 
shadows of Sgr~A* and M87.

\section{The Pleba{\'n}ski--Demia{\'n}ski metrics}
\label{sec:PD}
We consider the general Pleba{\'n}ski--Demia{\'n}ski class of stationary, 
axially symmetric type D solutions of the Einstein--Maxwell equations
with a cosmological constant. 
As a matter of fact, these solutions were first found by Debever \cite{Debever.1971} 
in 1971 but are better known in the form of Pleba{\'n}ski and Demia{\'n}ski 
\cite{PlebanskiDemianski.1976} from 1976. 
The Pleba{\'n}ski--Demia{\'n}ski metrics are discussed in detail  
by Griffiths and Podolsk{\'y}\cite{GriffithsPodolsky.2009} and by 
Stephani et al. \cite{StephaniKramer.2003}. 
It is common to use rescaled units, i.e. units so that the speed 
of light and the gravitational constant are normalized ($c=1$, $G=1$). 
In Boyer--Lindquist coordinates this metric is then given by, 
see Ref.~\citen{GriffithsPodolsky.2009}, p.~311
\begin{multline}
	g_{\mu \nu} \drm x^{\mu} \drm x^{\nu} = \frac{1}{\Omega^{2}} \biggl(
		\Sigma \bigl( \tfrac{1}{\Delta_{r}}\drm r^2 
		+ \tfrac{1}{\Delta_{\vartheta}}\drm \vartheta^2 \bigr) 
	+ \frac{1}{\Sigma} 
		\Bigl( (\Sigma + a\chi)^2 \Delta_{\vartheta}\sin^2\vartheta 
			- \Delta_{r} \chi^2 \Bigr) \drm \varphi^2 \\
	+ \frac{2}{\Sigma} 
		\Bigl( \Delta_{r}\chi - a(\Sigma + a\chi) 
			\Delta_{\vartheta}\sin^2\vartheta \Bigr) \drm t \drm \varphi 
	- \frac{1}{\Sigma} 
		\Bigl( \Delta_{r} - a^2\Delta_{\vartheta}\sin^2\vartheta \Bigr) \drm t^2
	\biggr)
	\label{eq:Metric}
\end{multline}
with the abbreviations 
\begin{align}
&\begin{aligned}
	\Omega &= 1 - \tfrac{\alpha}{\omega} (\ell + a \cos\vartheta) r, \\
	\Sigma &= r^2 + (\ell + a\cos\vartheta)^2, \\
	\chi   &= a\sin^2\vartheta - 2\ell(\cos\vartheta + C),
\end{aligned}
&
&\begin{aligned}
	\Delta_{\vartheta} &= 1 - a_{3}\cos\vartheta - a_{4}\cos^{2}\vartheta, \\
	\Delta_{r} &= b_{0} + b_{1}r + b_{2}r^{2} + b_{3}r^{3} + b_{4}r^{4}.
\end{aligned}
	\label{eq:MetricFunc}
\end{align}
The coefficients of the polynomials $\Delta_{\vartheta}$ and $\Delta_{r}$ are
\begin{align}
&
\label{eq:MetricFunc_ai}
\begin{aligned}
	a_{3} &= 2\tfrac{\alpha}{\omega}am - 4a\ell\bigl(
		\tfrac{\alpha^{2}}{\omega^{2}}(k+\beta) + \tfrac{\Lambda}{3} \bigr),\\ 
	a_{4} &= -a^{2} \bigl(
		\tfrac{\alpha^{2}}{\omega^{2}}(k+\beta) + \tfrac{\Lambda}{3} \bigr),
\end{aligned} \\[\medskipamount]
&
\label{eq:MetricFunc_bi}
\begin{aligned}
	b_{0} &= k+\beta, \\
	b_{1} &= -2m, \\
	b_{2} &= \tfrac{k}{a^{2}-\ell^{2}} + 4\tfrac{\alpha}{\omega}\ell m 
		- (a^{2}+3\ell^{2})\bigl(
		\tfrac{\alpha^{2}}{\omega^{2}}(k+\beta) + \tfrac{\Lambda}{3} \bigr), \\
	b_{3} &= -2\tfrac{\alpha}{\omega} \Bigl( \tfrac{k\ell}{a^{2}-\ell^{2}} 
		- (a^{2}-\ell^{2}) \Bigl( \tfrac{\alpha}{\omega}m - \ell \bigl(
		\tfrac{\alpha^{2}}{\omega^{2}}(k+\beta) + \tfrac{\Lambda}{3} \bigr) 
		\Bigr) \Bigr), \\
	b_{4} &= -\bigl(\tfrac{\alpha^{2}}{\omega^{2}}k + \tfrac{\Lambda}{3} \bigr)
\end{aligned}
\end{align}
with
\begin{align}
	k &= \frac{1 + 2\tfrac{\alpha}{\omega}\ell m - 3\ell^{2}\bigl( 
		\tfrac{\alpha^{2}}{\omega^{2}}\beta + \tfrac{\Lambda}{3} \bigr)}{
		1 + 3\tfrac{\alpha^{2}}{\omega^{2}}\ell^{2}(a^{2}-\ell^{2})}
		(a^{2}-\ell^{2}), &
	\omega &= \sqrt{a^{2}+\ell^{2}}.
	\label{eq:MetricFunc_k_omega}
\end{align}

Basically, the coordinates $t$ and $r$ may range over all of $\mathbb{R}$
while $\vartheta$ and $\varphi$ are standard coordinates on the two-sphere. 
Note, however, that for some values of the black-hole parameters $r$
and $\vartheta$ have to be restricted, see below. 
The Pleba{\'n}ski--Demia{\'n}ski space-time depends on seven parameters, 
$m$, $a$, $\beta$, $\ell$, $\alpha$, $\Lambda$ and $C$, which are to be 
interpreted in the following way. $m$ is the mass of the black hole and 
$a$ is its spin. $\beta$ is a parameter that comprises electric and magnetic 
charge, $\beta=q_{e}^{2}+q_{m}^{2}$, if non-negative; if $\beta$ is negative, 
the metric cannot be interpreted as a solution to the Einstein--Maxwell 
equations but metrics of this form occur in some brane-world scenarios
\cite{AlievGumrukcuoglu.2005}. The NUT parameter $\ell$ is to be interpreted 
as a gravitomagnetic charge.
The parameter $\alpha$ gives the acceleration of the black hole which is at 
the center of our investigation while $\Lambda$ is the cosmological constant. 
The parameter $C$, which was introduced by Manko and Ruiz\cite{MankoRuiz.2005}, 
is relevant only if $\ell\neq 0$. In this case there is a (conic) singularity 
somewhere on the $z$ axis and by choosing $C$ appropriately this singularity 
can be distributed symmetrically or asymmetrically on the positive and the 
negative $z$ axis. Note that this parameter $C$ has nothing to do with the 
name \blqq $C$-metric\brqq\ for the accelerated Schwarzschild space-time.
All the parameters, $m$, $a$, $\ell$, $\beta$, $\Lambda$, $\alpha$ and $C$, 
may take arbitrary real values in principle, albeit not all possibilities 
are physically relevant. 

If $\alpha=0$, the Pleba{\'n}ski--Demia{\'n}ski 
class reduces to the Pleba{\'n}ski space-times\cite{Plebanski.1975}
which are also known as the Kerr--Newman--NUT--(anti-)de Sitter 
space-times. For this more special class of metrics the photon regions 
and the shadows have been discussed in our earlier paper, see 
Ref.~\citen{GrenzebachPerlick.2014}. The Pleba{\'n}ski--Demia{\'n}ski class 
covers many well-known non-accelerated ($\alpha=0$) space-times like the 
Schwarzschild ($a=\beta=\ell=\Lambda=0$), Kerr ($\beta=\ell=\Lambda=0$), 
or Reissner--Nordstr{\"o}m space-time ($a=\ell=\Lambda=0$) as well as 
the accelerated $C$-metric ($a=\beta=\ell=\Lambda=0$) or their rotating version
($\beta=\ell=\Lambda=0$) which we will call \emph{accelerated Kerr space-time}.

The metric \eqref{eq:Metric} becomes singular at the roots of $\Omega$, 
$\Sigma$, $\Delta_{r}$, $\Delta_{\vartheta}$ and $\sin\vartheta$.
Some of them are mere coordinate singularities while others are
true (curvature) singularities. In the following we briefly discuss 
the influence of $\alpha$ on these singularities.

\begin{figure}[bpt]
	\includegraphics[width=\textwidth]{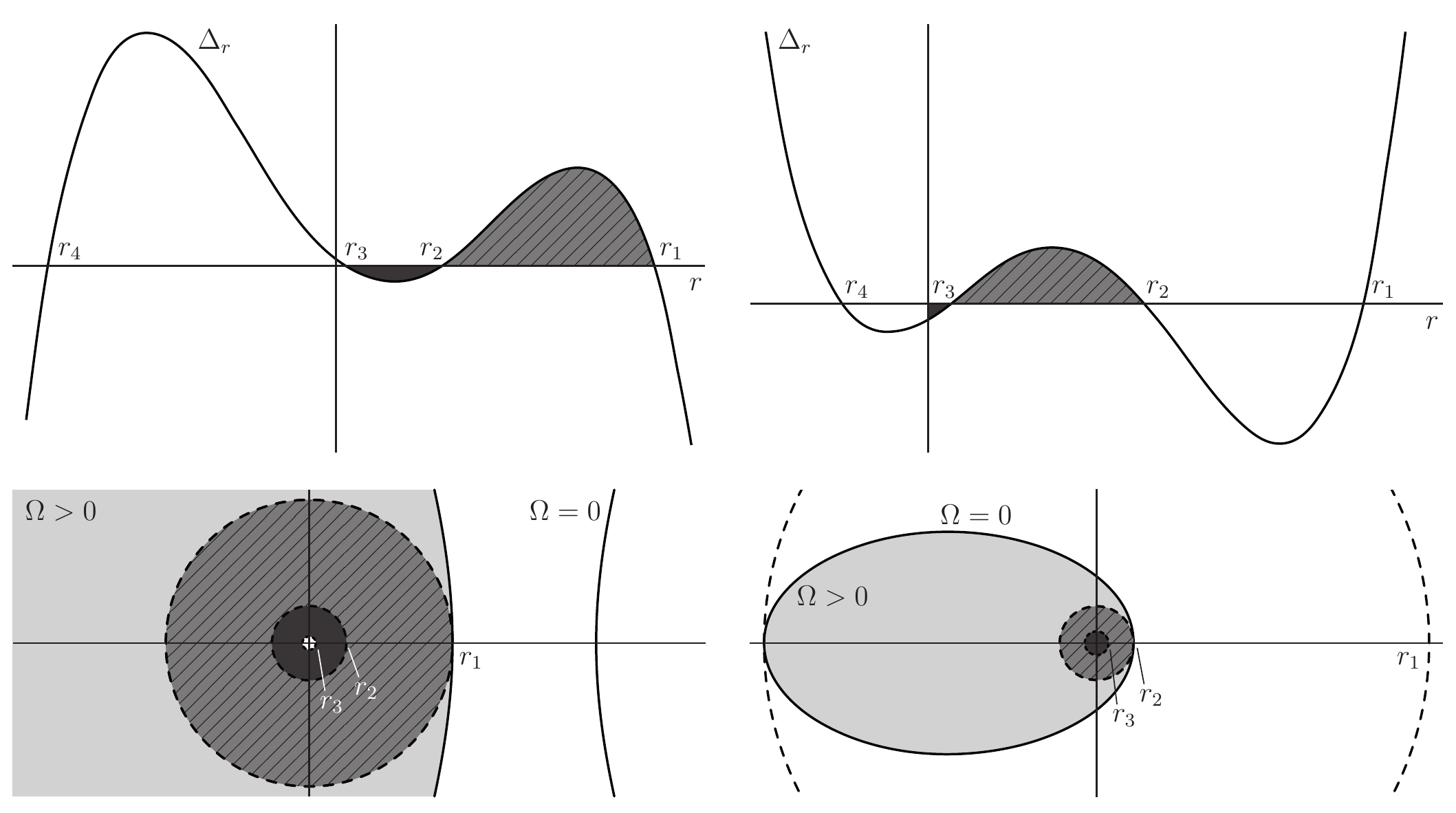}
	\caption{A schematic illustration of the graph of $\Delta_r$ (upper row)
	and a polar plot of the region $\Omega>0 $ (lower row) given 
	by \eqref{eq:OmegaSing}. Depending on the sign of the leading 
	coefficient $b_4$, see \eqref{eq:MetricFunc_bi}, $\Delta_r$ goes to 
	$+\infty$ of $- \infty$ for big radii $r$; 
	the space-times (with $\Lambda\geq 0$) belonging to the plots in the 
	left column are dominated by the Kerr property ($\ell^2 < a^2$) and in the 
	right column by the NUT property ($a^2 < \ell^2$).
	The space-time is restricted to that region where $\Omega>0$
	which is marked here with a light gray shading. Geometrically, the
	boundary of this region is one sheet of a two-sheeted hyperboloid 
	(left) or an ellipsoid (right).
	As in the NUT dominated case (right) the root $r_1$ of $\Delta_r$
	is not inside the allowed ellipsoid $\Omega>0$, it could be no event 
	horizon. Thus, the event horizon of the black hole is instead at 
	$r_3$ (right) or at $r_2$ (left). The gray-shaded and hatched region 
	marks the outer domain of communication ($\Delta_r>0$) where we will 
	place the observers.
	}
\label{fig:horizons}
\end{figure}

$\Omega$ becomes zero if 
\begin{equation}
	r = \frac{\sqrt{a^{2}+\ell^{2}}}{\alpha(\ell + a \cos\vartheta)}.
\label{eq:OmegaSing}
\end{equation}
As the metric blows up if $\Omega \to 0$, Eq. (\ref{eq:OmegaSing})
determines the boundary of the space-time, i.e., we have to restrict
to the region where $\Omega$ is positive. The allowed region is a half-space 
bounded by a plane ($\ell=0$), a half-space bounded by one sheet of a 
two-sheeted hyperboloid ($\ell^2 < a^2$), a domain bounded by a 
cylinder ($\ell^2 = a^2$), or a domain bounded by an ellipsoid ($\ell^2 > a^2$), 
see Fig.~\ref{fig:horizons}. For $\alpha=0$ there is no restriction 
because $\Omega\equiv 1$. 

$\Sigma$ becomes zero at the ring singularity
\begin{equation}
	r=0 \quad \text{and} \quad \cos\vartheta = -\ell/a.
\label{eq:RingSing}
\end{equation}
It exists for $\ell^2 < a^2$ and is a curvature singularity (if $m \neq 0$). 
Outside of this singularity the sphere $r=0$ is regular, so it is possible 
to travel through one of these two hemispheres (``throats'') from the 
region $r>0$ to the region $r<0$ and vice versa. If 
$\ell^2 > a^2$, there is no ring singularity and the sphere $r=0$ is 
regular everywhere. In the limiting case where $\ell^2 = a^2$ the ring 
singularity degenerates into a point on the axis. It becomes a point 
singularity for $\ell = a = 0$ that disconnects the space-time into the 
regions $r>0$ and $r<0$. The ring singularity is unaffected by $\alpha$.

Moreover, the metric is singular on the $z$ axis, i.e. if $\sin \vartheta 
=0$. If $\alpha\neq0$ oder $\ell\neq0$, this is a true (conical) singularity 
on (at least a part of) the rotational axis. In the NUT case the singularity 
depends on the Manko-Ruiz parameter $C$.
For further details we refer to the book by Griffiths and 
Podolsk{\'y}\cite{GriffithsPodolsky.2009}.

The real roots of $\Delta_{r}$ yield coordinate singularities which 
correspond to the up to $4$ horizons $r_1 > r_2 > \ldots$ of the space-time. 
If $\alpha=0$ and $\Lambda=0$, then $\Delta_{r}$ reduces to a second-degree
polynomial, $\Delta_{r} = r^2 - 2mr + a^2 - \ell^2 + \beta$, and horizons 
can be found at 
\begin{equation}
	r_{\pm} = m \pm \sqrt{m^2 -a^2 +\ell^2 - \beta}
	\label{eq:InOutHorizon}
\end{equation}
if $a^{2} \le a_{\mathrm{max}}^{2} := m^{2}+\ell^{2}-\beta$; 
then $r_{+}(=r_1)$ is the outer (event) horizon of the black hole and 
$r_{-}(=r_2)$ is the inner horizon. The \emph{domain of outer 
communication} is the region outside of the outer horizon where 
$\Delta_r>0$. For $a^{2} > a_{\mathrm{max}}^{2}$ we would find,
instead of a black hole, a naked singularity or a regular space-time. 
Since we are interested only in the black hole case, we will not
consider this possibility in the following.
In the accelerated or cosmological scenario ($\alpha\neq0$ or $\Lambda\neq0$)
the horizons could not in general be specified in a simple form because 
$\Delta_{r}$ is then a polynomial of degree $4$. 
Depending on the sign of the leading coefficient $b_4$, which is mostly 
determined by whether $a^2>\ell^2$ and by the sign of $\Lambda$, the 
vector field $\partial_r$ is timelike or spacelike for big values of $r$, 
see Fig.~\ref{fig:horizons}. 
In the timelike case (left column in Fig.~\ref{fig:horizons}), all real
roots of $\Delta_r$ are in the allowed region with $\Omega > 0$. Hence, 
the first root $r_1$ represents a cosmological horizon and the 
subsequent root $r_2$ is the black-hole horizon. In this case the 
domain of outer communication is the region between $r_1$ and $r_2$ where
$\Delta_r>0$ (gray shaded and hatched region in Fig.~\ref{fig:horizons}). 
In the case that $\partial _r$ is spacelike for big $r$ (right column 
in Fig.~\ref{fig:horizons}), the first root $r_1$ is not in the allowed 
region. Hence, we have at $r_2$ a cosmological horizon, at $r_3$ the 
event horizon of the black hole, and in between the outer domain of 
communication.
The horizons can be easily determined if $\beta=\ell=\Lambda =0$.
Then $k = a^{2}$ and $\omega = a$, hence
\begin{equation}
	\Delta_{r} = (r^2 - 2mr + a^2) (1 - \alpha^{2}r^{2})
	\label{eq:Dr0}
\end{equation}
and we find the usual (Kerr) horizons at $r=r_\pm$ given by 
\eqref{eq:InOutHorizon} with $\ell =0$ and $\beta =0$, and the 
additional horizons at $r=\pm\frac{1}{\alpha}$. Of course, we 
must have $|\alpha| < \tfrac{1}{r_+}$.

In general, at the roots of $\Delta_\vartheta$ would be coordinate 
singularities, too; these would indicate further horizons where the 
vector field $\partial_{\vartheta}$ would change the causal 
character from spacelike to timelike, just as the vector field $\partial_r$ 
does at the roots of $\Delta_r$. However, since these horizons would lie 
on cones $\vartheta = \mathrm{constant}$ instead of on spheres 
$r = \mathrm{constant}$ such a situation would be hardly of any physical 
relevance. Therefore, we exclude it by limiting the parameters of 
the black hole appropriately. As $\Delta_\vartheta =0$ implies
\begin{equation}
	\cos\vartheta_{\pm} = \frac{-a_{3} \pm \sqrt{a_{3}^{2} + 4a_{4}}}{2a_{4}},
	\label{eq:ConicSing}
\end{equation}
$\Delta_{\vartheta} \neq 0$ is guaranteed for all real $\vartheta$ 
if the radicand in (\ref{eq:ConicSing}) is negative. Therefore, we 
agree to choose the black-hole parameters such that $0>a_{3}^{2} + 4a_{4}$.  
If $\beta=\ell=\Lambda =0$, this condition can be simplified.
Then $k = a^{2}$ and $\omega = a$, hence
\begin{equation}
	\Delta_{\vartheta} = 1 - 2\alpha m \cos\vartheta + \alpha^{2}
   a^{2}\cos^{2}\vartheta, 
   \label{eq:Dtheta0}
\end{equation}
and $\Delta _{\vartheta} \neq 0$ is assured if 
\begin{equation}
	| \alpha | < 
	\begin{cases}
		\frac{1}{2m} & 
			\text{if $a=0$},\\
		\frac{r_{-}}{a^{2}}=\frac{m-\sqrt{m^{2}-a^{2}}}{a^{2}} & 
			\text{if $a\neq0$} 
	\end{cases}
\end{equation}

There are some other interesting regions around a black hole characterized by 
the change of the causal character of the Killing vector fields $\partial_t$ 
and $\partial_{\varphi}$.

In that region where $\partial_t$ becomes spacelike, i.e. $g_{tt}>0$, no 
observer can move on a $t$-line. Thus, any observer in this 
region has to rotate (in $\varphi$ direction). This region with 
$g_{tt}>0$ is known as the \emph{ergosphere} or the   
\emph{ergoregion}\footnote{Some authors call only the intersection of the 
region where $g_{tt}>0$ with the domain of outer communication the 
\emph{ergoregion}. This is that part of the region $g_{tt}>0$ which an 
outside observer would be able to see.}. %
An ergoregion only exists if $a\neq 0$.
Note that at the horizons, i.e. at the roots of $\Delta_r$, the metric 
coefficient $g_{tt}= -\frac{1}{\Omega^{2}\Sigma} \bigl(\Delta_{r} 
- a^2\Delta_{\vartheta}\sin^2\vartheta \bigr)$ is positive. 
Hence, the horizons are always contained within the ergoregion. 
For $\alpha\neq0$ or $\Lambda\neq0$ there are ``cosmological'' horizons
in addition to the black-hole horizons; then the ergoregion consists 
of several connected components.
The boundary of (a component of) the ergoregion is always tangential to 
the horizon on the rotational axis, i.\,e. at $\vartheta=0, \pi$.

If $a \neq 0$ or $\ell \neq 0$, there are regions where the Killing 
field $\partial_{\varphi}$ becomes timelike, $g_{\varphi \varphi} =0$. 
This indicates causality violation, because the $\varphi$-lines are 
closed timelike curves. For $\ell=0$ the region where $g_{\varphi \varphi} =0$
is completely contained in the domain where $r<0$ and, thus, hidden 
behind the horizon for an observer in the domain of outer communication. 
In the case $\ell \neq 0$, however, there is a causality violating 
region in the domain of outer communication around the axial singularity. 

In the following, we will only consider the black-hole case where we have 
at least one positive root of $\Delta_r$. Observers will be placed in the 
domain of outer communication.

\section{Photon Regions}
\label{sec:regionK}
In the Pleba{\'n}ski class of space-times, i.e., for $\alpha = 0$,
the geodesic equation is completely integrable; in addition to the
obvious constants of motion, there is a fourth constant of motion, 
known as the Carter constant, which is associated with a second-rank
Killing tensor. If $\alpha \neq 0$, instead of this Killing tensor
we only have a conformal Killing tensor. This is sufficient to 
assure complete integrability for \emph{lightlike} geodesics. The
four constants of motion are the Lagrangian
\begin{align}
	\Lag &= \tfrac{1}{2} g_{\mu\nu}\dot{x}^{\mu}\dot{x}^{\nu},  &
	\Lag &= 0 \quad \text{for light},
	\label{eq:L}
\end{align}
the energy $E$ and the $z$-component $L_z$ of the angular momentum
\begin{align}
	E :&= -\frac{\partial\Lag}{\partial\dot{t}} 
		= -g_{\varphi t}\dot{\varphi} -g_{tt}\dot{t},
	&
	L_{z} :&= \frac{\partial\Lag}{\partial\dot{\varphi}}
		= g_{\varphi\varphi}\dot{\varphi} +g_{\varphi t}\dot{t},
	\label{eq:E_Lz}
\end{align}
and the Carter constant $K$, see Ref.~\citen{Carter.1968b}, which is 
now associated only with a conformal Killing tensor. 
The Carter constant may be viewed as the separation constant 
for the $r$ and the $\vartheta$ motion of lightlike geodesics. %
The four constants of motion allow us to write the lightlike
geodesic equation in separated first-order form, 
\begin{subequations}\label{eq:EoM}
\begin{align}
	\frac{\Sigma}{\Omega^{2}} \dot{t}
	&= \frac{\chi(L_{z}-E\chi)}{\Delta_{\vartheta} \sin^{2}\vartheta}
		+\frac{(\Sigma + a\chi) \bigl((\Sigma + a\chi)E - aL_{z}\bigr)}{\Delta_{r}},
\label{eq:EoM_t} \\
	\frac{\Sigma}{\Omega^{2}} \dot{\varphi}
	&= \frac{L_{z}-E\chi}{\Delta_{\vartheta} \sin^{2}\vartheta}
		+\frac{a\bigl((\Sigma + a\chi)E - aL_{z}\bigr)}{\Delta_{r}},
\label{eq:EoM_phi} \\
	\biggl(\frac{\Sigma}{\Omega^{2}}\biggr)^{2} \dot{\vartheta}^{2} 
	&= \Delta_{\vartheta}K - \frac{(\chi E - L_{z})^{2}}{\sin^{2}\vartheta} 
	=: \Theta(\vartheta),
\label{eq:EoM_theta} \\
	\biggl(\frac{\Sigma}{\Omega^{2}}\biggr)^{2} \dot{r}^{2} 
	&= \bigl((\Sigma + a\chi)E-aL_{z}\bigr)^{2} - \Delta_{r}K
	=: R(r).
\label{eq:EoM_r}
\end{align}
\end{subequations}
In order to derive an equation for the shadow of accelerated
black holes, we proceed in the same way as for the Pleba{\'n}ski
space-times. As  a first step, we have to determine the spherical 
lightlike geodesics, i.e., lightlike geodesics staying on a sphere
$r = \mathrm{constant}$. The region filled by these spherical geodesics is 
called the \emph{photon region} $\K$. Mathematically, spherical orbits 
are characterized by $\dot{r}=0$ and $\ddot{r}=0$ which requires by 
\eqref{eq:EoM_r} that $R(r)=0$ and $R'(r)=0$. Thus
\begin{align}
	K_{E} &= \frac{\bigl((\Sigma + a\chi)-aL_{E}\bigr)^{2}}{\Delta_{r}},  &
	K_{E} &= \frac{4r\bigl((\Sigma + a\chi)-aL_{E}\bigr)}{\Delta_{r}'},
\label{eq:2xKE}
\end{align}
where $\Delta_{r}'$ is the derivative of $\Delta _r$ with respect to $r$ 
and $K_{E}$, $L_{E}$ are abbreviations
\begin{align}
	K_{E} &= \frac{K}{E^{2}}, & L_{E} &= \frac{L_{z}}{E}.
\label{eq:LEKE}
\end{align}
After solving \eqref{eq:2xKE} for the constants of motion 
\begin{align}
	K_{E} &= \frac{16r^{2}\Delta_{r}}{(\Delta_{r}')^{2}}, &
	aL_{E} &= \bigl(\Sigma + a\chi\bigr) - \frac{4r\Delta_{r}}{\Delta_{r}'},
\label{eq:KL_SphLR}
\end{align}
we can substitute these expressions into \eqref{eq:EoM_theta}. As the 
left-hand side of \eqref{eq:EoM_theta} is non-negative, we find 
an inequality that determines the photon region
\begin{equation}
	\K\colon \bigl(4r\Delta_{r} - \Sigma \Delta_{r}' \bigr)^{2}
		\leq 16 a^{2} r^{2} \Delta_{r} \Delta_{\vartheta} \sin^{2}\vartheta.
\label{eq:regionK}
\end{equation}
Of course, the equality sign determines the boundary of the photon region.

Just as in the non-accelerated space-times\cite{Perlick.2004,GrenzebachPerlick.2014}
for every point ($r_{p}, \vartheta_{p}$) of $\K$ there is a lightlike 
geodesic through ($r_{p}, \vartheta_{p}$) that stays on the sphere $r=r_{p}$. 
The $\vartheta$ motion is an oscillation bounded by the boundary of $\K$
while the $\varphi$ motion given by \eqref{eq:EoM_phi} might be
rather complicated. 

The stability of these spherical geodesic with respect to radial perturbations 
is determined by the sign of $R''$; a spherical geodesic at $r=r_{p}$ is unstable
if $R''(r_{p})>0$, and stable if $R''(r_{p})<0$. From \eqref{eq:EoM_r} 
we get with \eqref{eq:KL_SphLR}
\begin{equation}
	\frac{R''(r)}{8 E^{2}} \Delta_{r}'^{2} = 
		2r\Delta_{r}\Delta_{r}' + r^{2} \Delta_{r}'^{2} 
		- 2r^{2} \Delta_{r} \Delta_{r}''.
	\label{eq:stability}
\end{equation}

A non-rotating black hole ($a=0$) is surrounded by a \emph{photon sphere}, 
rather than  by a photon region, since the inequality \eqref{eq:regionK} 
defining $\K$ reduces to an equality
\begin{equation}
	4r\Delta_{r} = (r^2+\ell ^2 ) \Delta_{r}'.
\label{eq:ps}
\end{equation}
The best known example is the photon sphere at $r=3m$ in the Schwarzschild 
space-time.

\begin{figure}[htbp]
	\centering
	\includegraphics{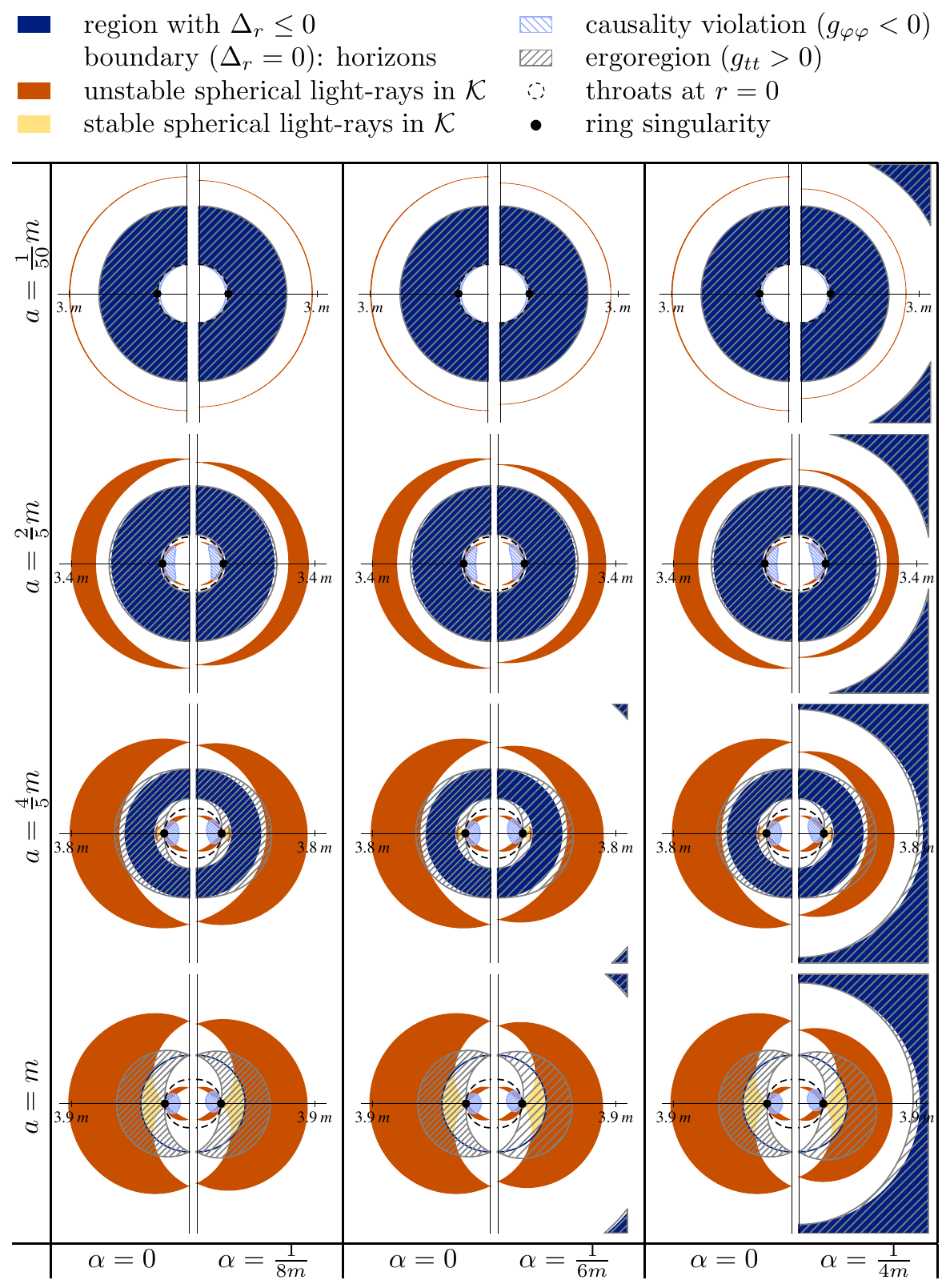}
	\caption[Photon regions in accelerated Kerr space-time.]{%
	Photon regions in accelerated Kerr space-time for spins 
	$a=\lambda a_{\mathrm{max}}$, where $a_{\mathrm{max}}=m$. 
	In each column the plot for the unaccelerated Kerr space-time (left) 
	is compared to the plot for an accelerated Kerr space-time (right). %
	The specific acceleration parameters are listed in the bottom row. %
	A legend for the plotted regions is given at the top.}
	\label{fig:Photonregion}
\end{figure}

Because of the rotational symmetry it is convenient to plot a meridional 
section through space-time for illustrating the regions around a black hole.
The resulting pictures, which are ($r$,$\vartheta$) polar diagrams where 
$\vartheta$ is measured from the positive $z$-axis, are shown in 
Fig.~\ref{fig:Photonregion}. Each figure contains the photon region $\K$, 
where unstable and stable spherical light rays are distinguished according 
to \eqref{eq:stability}, the horizons $r_\pm$ of the black hole
given as boundaries of the region where $\Delta_{r}\leq 0$,
the ergoregion, the causality violating region, and the ring singularity.

The dashed circle marks the throats at the sphere $r=0$.
For viewing the whole range of the space-time, we use two different scales
for the radial coordinate:
In the inner region $r<0$ (inside the sphere $r=0$) the radial coordinate 
is plotted as $m \exp \big( r/m \big)$; this is continuously extended with 
$r+m$ in the outer region $r>0$ (outside the sphere $r=0$). 
By not plotting just the exponential of the Boyer--Lindquist coordinate $r$, 
as suggested by O'Neill\cite{ONeill.1995}, we avoid a strong deformation of 
the outer part.

While our formulas apply to black holes of the entire 
Pleba{\'n}ski--Demia{\'n}ski class, in the pictures we restrict to 
accelerated Kerr space-times ($\beta = \ell = \Lambda = 0$) because 
we want to focus on the effect of the acceleration.
Fig.~\ref{fig:Photonregion} comprises various images of photon regions
$\K$ for different spin and acceleration parameters, 
where the spin is varied in the rows and the acceleration in the columns. 
For the spin we choose fractions of the value for an extremal black 
hole $a=\lambda a_{\mathrm{max}}$ with
$\lambda\in\bigl\{\frac{1}{50}, \frac{2}{5}, \frac{4}{5}, 1\bigr\}$
and $a_{\mathrm{max}}=m$, cf.~\eqref{eq:InOutHorizon} and \eqref{eq:Dr0}.
Although one would expect only very small acceleration parameters in reality, 
we choose relatively big values ($\alpha\in\bigl\{0, \frac{1}{8m}, 
\frac{1}{6m}, \frac{1}{4m}\bigl\}$) for a better illustration of the effects.
For each of the three values of the acceleration the figure is
compared with the ordinary Kerr case. 

In the ordinary Kerr space-time ($\alpha=0$), see the left half of the images 
in the columns of Fig.~\ref{fig:Photonregion}, we see two photon regions: 
one with unstable orbits in the exterior region of the black hole at $r>r_+$ 
and one in the interior region at $r<r_-$ which contains unstable orbits as
well as stable ones. %
For spinning black holes the exterior photon region develops a 
crescent-shaped cross-section which grows with increasing spin $a$. %
The inner photon region consists of two parts divided by the ring singularity. %
Note that also in the rotating case there are circular photon orbits, 
namely at that five points on the boundary of $\K$ which are tangent to a 
sphere $r=\mathrm{constant}$: there are three circular photon orbits
in the equatorial plane---two at the boundary of the exterior photon region 
and one at the boundary of the interior photon region---and two more
off the equatorial plane at the boundary of the interior photon region 
where $r<0$. %
Furthermore, we find the ergoregion containing the horizons of the black
hole, and in the interior adjacent to the ring singularity a 
causality violating region. If $a^2 > m^2/2$ the ergoregion intersects
the exterior photon region. %
All of these regions are symmetric with respect to the equatorial plane. 

\begin{figure}[htbp]
	\centering
	\includegraphics{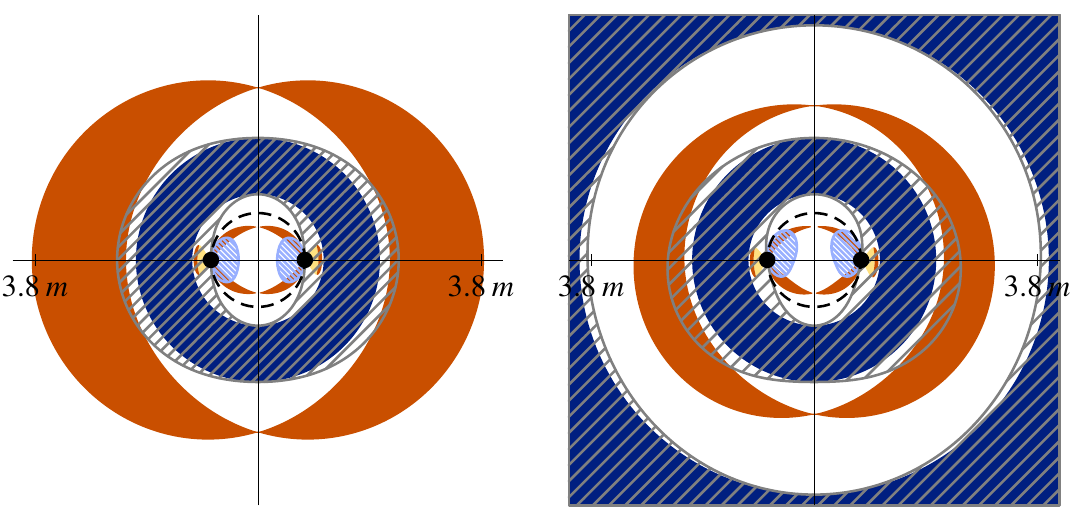}
	\caption[Photon regions in accelerated Kerr space-time.]{%
	Bigger versions of both plots shown in the fourth row of the third 
	column in Fig.~\ref{fig:Photonregion}. %
	Illustrated are the photon regions of spinning Kerr black holes 
	($a=\frac{4}{5}m$) where the left plot belongs to the ordinary 
	space-time ($\alpha=0$) and the right plot to an accelerated 
	space-time ($\alpha=\frac{1}{4m}$).}
	\label{fig:Photonregion2}
\end{figure}

The plots for the accelerated Kerr space-times look similar to the 
non-accelerated ones but there are two significant differences. Firstly, 
a non-zero acceleration parameter gives rise to additional horizons,
similarly to a cosmological constant. Secondly, the plots are no longer 
symmetric with respect to the equatorial plane which is similar to the
NUT case. The additional outer horizon, a cosmological one, is best seen 
in the illustration for the highest acceleration $\alpha=\frac{1}{4m}$. 
In principle, such a horizon also appears in all other plots but most or 
even all of it is located outside of the shown clipping. %
The asymmetry with respect to the equatorial plane is  best seen for 
$\alpha=\frac{1}{4m}$. With the exception of the causality violating region, 
the entire picture looks as if pushed into the negative $z$ direction, i.e., 
into the direction against the direction of the acceleration. For a better 
view, Fig.~\ref{fig:Photonregion2} shows bigger versions of the two plots 
shown in the fourth row of the third column in Fig.~\ref{fig:Photonregion}.

As one would expect, the photon region, the ergoregion and the causality 
violating region depend on the signs of $a$ and $\alpha$. While the
photon region is reflected at the equatorial plane if the sign of 
$\alpha$ is changed, the ergoregion and the causality violating region are 
reflected if the sign of $a\alpha$ is changed.
The effects of $\beta$, $\ell$ and $\Lambda$ on the photon regions
have been discussed in our earlier paper, see 
Ref.~\citen{GrenzebachPerlick.2014}. We do not repeat this here
because there are no new qualitative aspects if $\alpha$ is present.

\section{Shadows of Black Holes}
\label{sec:shadow}
If one looks into the direction of a black hole then there is a region
on the sky which stays dark, provided that there are no light sources
between the observer and the black hole. This dark region is called
the \emph{shadow} of the black hole. 
To determine the shape of the shadow we consider light rays
which are sent into the \emph{past} from the position $(r_O, \vartheta_O)$ of a 
fixed\footnote{Because of the symmetry, it is enough to specify the $r$ and 
$\vartheta$ coordinate to define a fixed position in space-time.} observer
in the domain of outer communication.
Then we can distinguish between two types of lightlike geodesics: Those 
where the radial coordinate increases after possibly passing through
a minimum and those where the radial coordinate decreases until reaching 
the horizon at $r=r_+$. %
If we assume that there are light sources distributed in the universe,
but not between the observer and the black hole, geodesics of the first 
kind could reach a light source; so we assign brightness to the initial
direction of such a light ray. Correspondingly, we assign darkness
to the initial directions of light rays of the second kind, i.e., 
these initial directions determine the shadow of the black hole. %
The boundary of the shadow corresponds to light rays on the borderline
between the two kinds. These are light rays that spiral asymptotically 
towards one of the unstable spherical light orbits in the exterior photon 
region $\K$.
Hence, the essential information for determining the shadow of a black hole
is in the surrounding photon region. One may even say that the shadow 
is an image of the photon region (but not of the event horizon).

For deriving an analytical formula for the boundary curve of the 
shadow we proceed, again, as in the case without acceleration.
First, we choose an orthonormal tetrad, cf. page 307 in 
Ref.~\citen{GriffithsPodolsky.2009}, for our fixed observer at
$(r_{O},\vartheta_{O})$
\begin{equation}
\begin{aligned}
	e_{0} &= \Omega \left.
		\frac{(\Sigma + a \chi) \partial_t + a \partial_{\varphi}}{
		\sqrt{\Sigma \Delta_r}}\right|_{(r_O,\vartheta_O)}, \quad &
	e_{2} &= -\Omega \left.
		\frac{(\partial_{\varphi} + \chi \partial_t)}{
		\sqrt{\Sigma \Delta_{\vartheta}}  \sin \vartheta} 
		\right|_{(r_O,\vartheta_O)}, \\[\smallskipamount]
	e_{1} &= \Omega \left.
		\sqrt{\dfrac{\Delta _{\vartheta}}{\Sigma}} \, \partial_{\vartheta}
		\right|_{(r_O,\vartheta_O)}, &
	e_{3} &= -\Omega \left.
		\sqrt{\frac{\Delta_r}{\Sigma}} \, \partial_r
		\right|_{(r_O,\vartheta_O)}.
\end{aligned}
\label{eq:newcoord}
\end{equation}
Since our observer is in the domain of outer communication, $\Delta_r$ is 
positive. $\Sigma$ is positive everywhere (except at the ring singularity 
which is not part of the space-time and, moreover, away from the domain of 
outer communication) and $\Delta_{\vartheta}$ is positive by assumption.
This guarantees real coefficients in Eqs.~\eqref{eq:newcoord}. It is easy
to check that the $e_i$ are orthonormal. %
As usual, the timelike vector $e_0$ is interpreted as the four-velocity 
of our observer. By our choice of the tetrad, $e_0 \pm e_3$ are tangential 
to the \emph{principal null congruences} of our metric; $e_3$ points into 
the spatial direction towards the center of the black hole. So we have
chosen the four-velocity of our observer adapted to the symmetries of the
space-time in the sense that $e_0$ is in the intersection of the 
$t$-$\varphi$-plane and the plane spanned by the two principal null
directions.

\begin{figure}[htbp]
	\centering
	\includegraphics[scale=0.5]{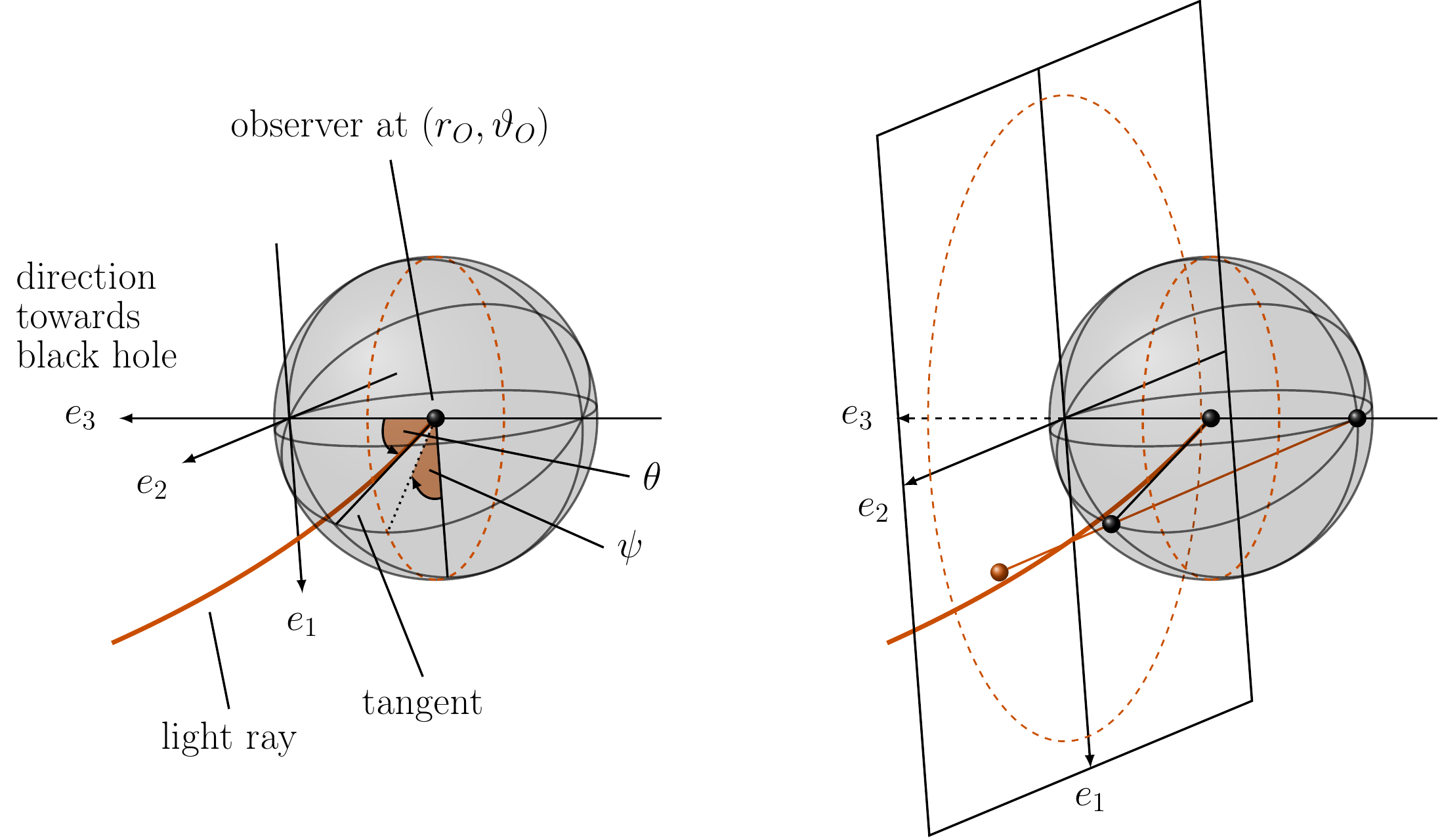}
	\caption{Eq. \eqref{eq:dotlambda2} defines celestial coordinates 
	$\theta$ and $\psi$ for the light rays at the observer's position,
	as illustrated in the left figure. With this choice, $\theta=0$ is 
	the direction towards the black hole.
	Every direction of a light ray represented by a point $(\theta, \psi)$
	on the celestial sphere (black ball) is visualized by its
	stereographic projection (red ball) on a plane. The 
	dotted (red) circles are the celestial equator $\theta = \pi/2$ 
	and its image under stereographic projection.}
	\label{fig:LightObserv}
\end{figure}

For any light ray $\lambda(s) = \bigl(r(s),\vartheta(s),\varphi(s),t(s)\bigr)$, 
the tangent vector at the position of the observer can be written in two 
different ways, using either the coordinate basis or the tetrad 
introduced above, 
\begin{align}
	\dot{\lambda} &= \dot{r} \partial_{r} + \dot{\vartheta} \partial_{\vartheta}
	+ \dot{\varphi} \partial_{\varphi} + \dot{t} \partial_{t},
\label{eq:dotlambda1} \\
	\dot{\lambda} &= \sigma \big( -e_{0} + \sin\theta \cos\psi e_{1} 
	+ \sin\theta \sin\psi e_{2} + \cos\theta e_{3} \big).
\label{eq:dotlambda2}
\end{align}
The second equation defines the celestial coordinates $\theta$ and 
$\psi$, see Fig.~\ref{fig:LightObserv}.
For the scalar factor $\sigma$ we obtain with \eqref{eq:E_Lz}
\begin{equation}
	\sigma = g\big(\dot{\lambda},e_0 \big) 
		= \Omega \left. 
		\frac{a L_z - (\Sigma+a\chi)E}{\sqrt{\Sigma \Delta_r}} 
		\right|_{(r_O,\vartheta_O)}.
\label{eq:sigma}
\end{equation}
We substitute $\dot{\varphi}$ and $\dot{r}$ from
\eqref{eq:EoM_phi} and \eqref{eq:EoM_r} into \eqref{eq:dotlambda1},
and we insert the expressions of $e_i$ from \eqref{eq:newcoord} 
into \eqref{eq:dotlambda2}. Then comparing the coefficients of
$\partial_{\varphi}$ and $\partial_r$ in the resulting two
equations yields
\begin{subequations}
\label{eq:thetapsi}
\begin{alignat}{3}
	T &:= & \sin\theta &= \left. 
		\frac{\sqrt{\Delta_r K_E}}{r^2 + \ell^2 - a\widetilde{L}_E} 
		\right|_{r=r_O},
	\label{eq:theta}
	\\[\smallskipamount]
	P &:= & \sin\psi &= \left.
		\frac{\widetilde{L}_E + a\cos^2\vartheta + 2\ell \cos\vartheta}{
			\sqrt{\Delta_{\vartheta} K_E} \sin\vartheta} 
		\right|_{\vartheta = \vartheta_O}, 
	\label{eq:psi}
\end{alignat}
\end{subequations}
where
\begin{equation}
	\widetilde{L}_E = L_E - a + 2 \ell C.
	\label{eq:tLE}
\end{equation}

If a light ray asymptotically approaches a spherical lightlike geodesic 
at a radius $r_p$ in the photon region, it must have the same constants 
of motion as this limiting spherical geodesic. By \eqref{eq:KL_SphLR} and 
\eqref{eq:tLE}, this implies that the constants of motion of light rays
that correspond to boundary points of the shadow are given by
\begin{align}
	K_{E}(r_p) &= \left. \frac{16r^{2}\Delta_{r}}{(\Delta_{r}')^{2}} \right|_{r=r_p}, 
	&
	a\widetilde{L}_{E}(r_p) &= \left. \Big(r^2 + \ell^2 - \frac{4r\Delta_{r}}{
		\Delta_{r}'} \Big)\right|_{r=r_p}.
\label{eq:LEKEc}
\end{align}
Here the range of $r_p$ is determined by the 
intersection of the exterior photon region \eqref{eq:regionK} with the cone 
$\vartheta = \vartheta_O$, cf. Ref.~\citen{GrenzebachPerlick.2014}. 
Thus, for a rotating black hole $r_p$ ranges over 
an interval at whose boundary points \eqref{eq:regionK} holds for 
$\vartheta = \vartheta_O$ with equality. 
If we insert \eqref{eq:LEKEc} into \eqref{eq:thetapsi}, we get the 
boundary curve of the shadow on the observer's sky parametrized with $r_p$.
In the case $a=0$ the photon region degenerates into a photon sphere
$r=r_p$. This unique value $r_p$ defines a unique $K_E(r_p)$ but 
does not restrict $\widetilde{L}_E$. Calculating the corresponding 
$\theta$ from \eqref{eq:thetapsi} gives the radius of the shadow which
is circular in this case. We may use $\widetilde{L}_E$ as a parameter
for the boundary curve, where $\widetilde{L}_E$ varies between the 
extremal values given by \eqref{eq:EoM_theta} 
for $\Theta(\vartheta_O)=0$.

Comparison with Ref.~\citen{GrenzebachPerlick.2014} shows that the 
formula \eqref{eq:regionK} for the photon region as well as the formulas 
(\ref{eq:thetapsi}, \ref{eq:tLE}, \ref{eq:LEKEc}) 
for the boundary curve of the shadow are identical with those of the 
non-accelerated case. However, the metric functions \eqref{eq:MetricFunc}
have now a more general meaning because they include the acceleration
parameter. %

Several properties of the shadow are preserved, even with the 
acceleration parameter added. A non-rotating black hole still 
has a circular shadow since \eqref{eq:theta} depends on the unique 
$K_E(r_p)$ but not on $\widetilde{L}_E$, so $\theta = \mathrm{constant}$
in this case. As the acceleration parameter breaks the spherical
symmetry, this is a non-trivial result. %
Furthermore, the shadow is still independent of the Manko-Ruiz parameter $C$
which is relevant only in the case $\ell \neq 0$. 
As in the non-accelerated case, the shadow is always symmetric with 
respect to a horizontal axis, because
$(\psi,\theta)$ and $(\pi-\psi, \theta)$ are determined by the same
constants of motion $K_E$ and $\widetilde{L}_E$. Again, this is a
non-trivial result because it is not implied by an underlying 
symmetry unless $\ell = 0$,
$\alpha = 0$ and $\vartheta _O = \pi /2$. 

It is to be emphasized that we have calculated the shape of the
shadow for an observer with a particular four-velocity, adapted to
the principal null directions of the space-time. For an observer
in a different state of motion, the shadow is distorted by 
aberration. These aberration effects have been discussed in
detail in Ref.~\citen{Grenzebach.2015}. As the aberration formula
maps circles onto circles, the statement that a non-rotating 
black hole produces a circular shadow is true for an observer
in \emph{any} state of motion.

\begin{figure}[tbp]
	\centering
	\includegraphics{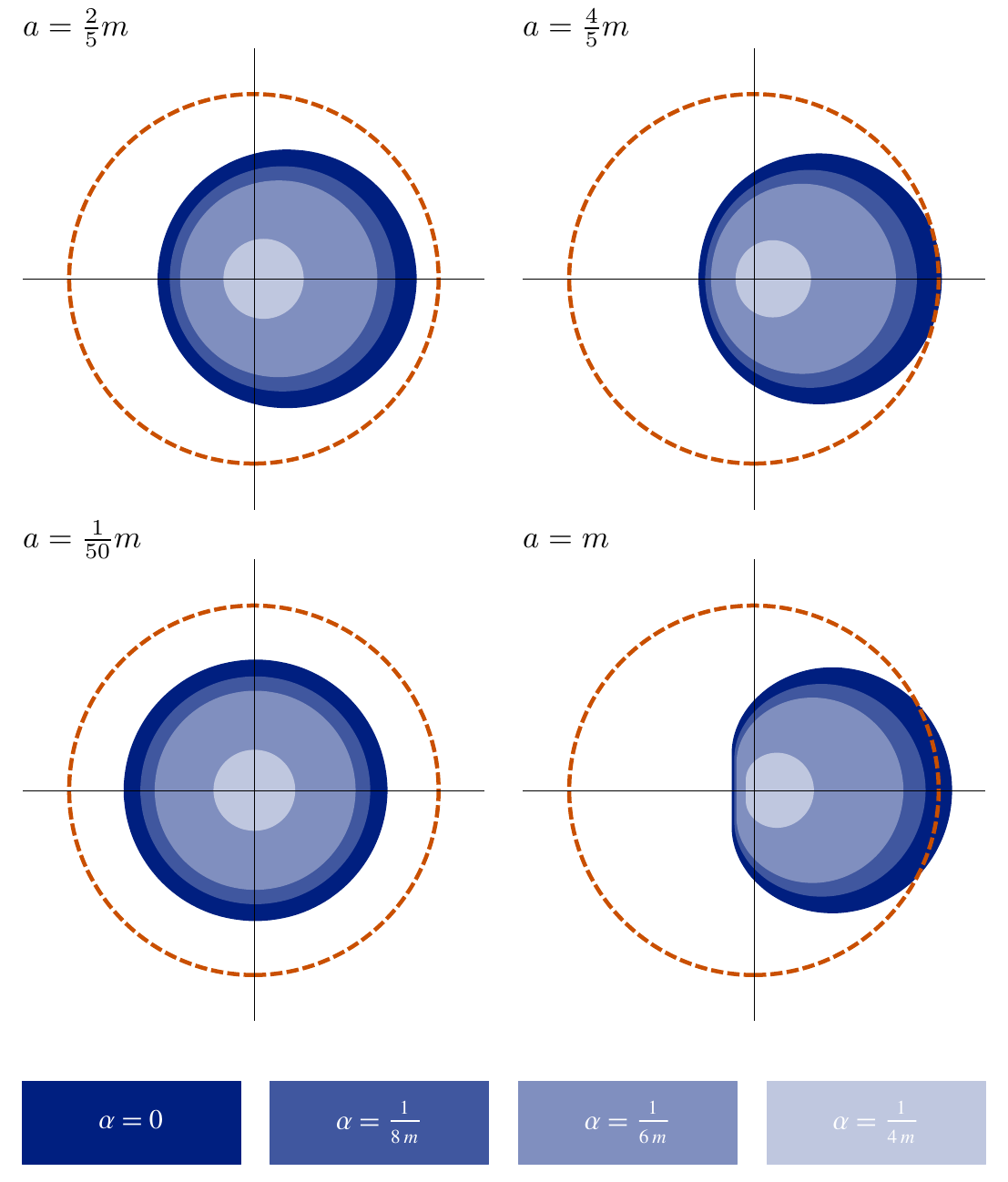}
	\caption[Shadows for accelerated Kerr space-times.]{Shadows of accelerated 
	Kerr black holes ($a_{\mathrm{max}}=m$) seen by an observer at $r_O=3.8 m$ 
	and $\vartheta_O = \pi /2$ for different spin values. The magnitude
	of the acceleration is color-coded where the specific values of	q
	$\alpha$ are listed below the plots.
	The dashed (red) circle marks the projection of the celestial equator, 
	cf. Fig.~\ref{fig:LightObserv}.}
	\label{fig:Shadow}
\end{figure}

\begin{figure}[tbp]
	\setlength{\tabcolsep}{5pt}
	\centering
	\includegraphics{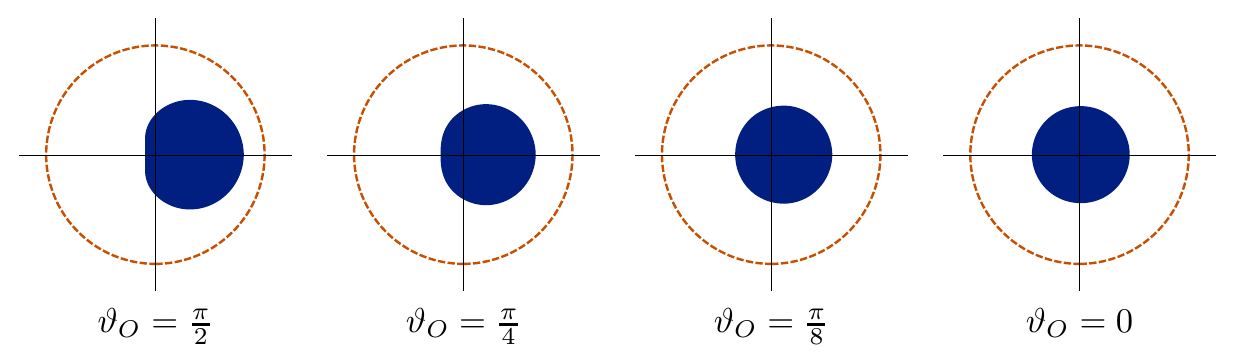}
	\caption[Plots of black hole's shadow for different observer 
		positions]{Shadow of a black hole in accelerated Kerr space-time 
		($\alpha=\frac{1}{6m}$, $a=m=a_{\mathrm{max}}$) for an observer at 
		$r_O = 3.8 m$ with different inclination angles $\vartheta_O$. 
		As in Fig.~\ref{fig:Shadow}, the dashed (red) circle indicates 
		the celestial equator.}
	\label{fig:ShadowObserv}
\end{figure}

Figures~\ref{fig:Shadow} and \ref{fig:ShadowObserv} comprise images of
shadows for different space-times seen by an observer at $r_O=3.8 m$ 
with varying inclination $\vartheta_O$. %
As explained in Fig.~\ref{fig:LightObserv}, we map the shadow onto a 
plane by stereographic projection.
Standard Cartesian coordinates in that plane of projection are given by
\begin{equation}
	\begin{pmatrix} 
		x(\rho) \\
		y(\rho)
	\end{pmatrix}
	= -2 \tan \big( \tfrac{1}{2}\theta(\rho) \big)
	\begin{pmatrix} 
		\sin \psi(\rho) \\
		\cos \psi(\rho) 
	\end{pmatrix}
\label{eq:stereo}
\end{equation}

In Fig.~\ref{fig:Shadow} we show the shadow for accelerated Kerr 
space-times where we have chosen the same values for $\alpha$ and $a$ 
as in Fig.~\ref{fig:Photonregion}. Here, the observer is fixed at 
Boyer--Lindquist coordinates $r_{O}=3.8m$ and $\vartheta_{O} = \pi/2$ 
(in the domain of outer communication). The different values of 
$\alpha$ are encoded into different shadings. 

Also with acceleration, the shape of the shadow is largely determined by 
the spin $a$ of the black hole. Hence, the shadow becomes more and more 
asymmetric with respect to a vertical axis with increasing spin $a$ where 
the asymmetry results from the \blqq dragging effect\brqq\ of the rotation 
on the light rays. 
The shadow is reflected at a vertical axis if the sign of $a$,
i.\,e. the spin direction, is changed. One might have expected a 
similar effect with respect to a horizontal axis if the sign of
$\alpha$ is changed. However, this is not true. As the
 shadow stays symmetric with respect to a  horizontal axis even 
if $\alpha\neq 0$, the shadow is independent of the direction 
of the acceleration, i.\,e. of the sign of $\alpha$. 
The acceleration has an effect on the \emph{size} of the
shadow, as is visible with the naked eye. This, however, has little
relevance in view of observations because the size also scales with 
$r_O$ and a comparison of the radius coordinates in different 
space-times has no direct operational meaning.

With the plots in Fig.~\ref{fig:ShadowObserv} we investigate the influence
of the observer's inclination $\vartheta _O$ on the shadow of an 
extremal Kerr black hole ($a=a_{\mathrm{max}}=m$) with acceleration
$\alpha=\frac{1}{6m}$.
As in Fig.~\ref{fig:Shadow} the observer ist fixed at $r_O=3.8m$. 
Clearly, for $\vartheta _O \to 0$ the shadow becomes circular.
We have already emphasized the remarkable fact that the shadow is 
always symmetric with respect to the horizontal axis.

\section{Angular Diameters of the Shadow of Black Holes}
\label{sec:angrad}
\begin{figure}[bp]
\floatbox[{\capbeside\thisfloatsetup{capbesideposition={right,bottom}}}]{figure}%
	{\caption{Angular radii of the shadow of a black hole. Owing to the symmetry 
	with respect to a horizontal axis, the two angular diameters (dashed lines) 
	of the shadow are given by three angular radii: two horizontal radii 
	$\varrho_{h_i}$ and one vertical radius $\varrho_{v}$. The angular 
	diameters are calculated as $\delta_h = \varrho_{h_1} + \varrho_{h_2}$ 
	and $\delta_v = 2\varrho_{v}$, respectively.}%
	\label{fig:AngRad}}%
	{\includegraphics[scale=1]{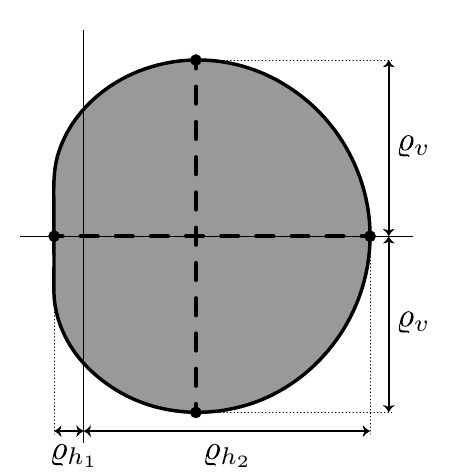}}
\end{figure}

From the analytical formulas \eqref{eq:thetapsi} and \eqref{eq:LEKEc} for 
the boundary curve of the shadow we can deduce expressions for the 
horizontal and vertical angular diameters of the shadow. 
These correspond to the dashed lines in Fig.~\ref{fig:AngRad}.
Owing to the symmetry, the angular diameters $\delta_h$ and $\delta_v$ are 
determined by three angular radii $\varrho_{h_1}$, $\varrho_{h_2}$, 
and $\varrho_{v}$ as indicated in Fig.~\ref{fig:AngRad}, 
\begin{alignat}{5}
	\delta_h &= \varrho_{h_1} + \varrho_{h_2}, & \qquad 
		\sin\varrho_{h_i} &= \sin\psi_{h_i}\sin\theta_{h_i} &&= P(r_{h_i}) T(r_{h_i}), 
\label{eq:AngDia_h} \\
	\delta_v &= 2\varrho_{v}, & \qquad 
		\sin\varrho_v &= \cos\psi_v \sin\theta)v &&= \sqrt{1-P^2(r_v)} T(r_v),
\label{eq:AngDia_v}
\end{alignat}
where $T$ and $P$ have the same meaning as in (\ref{eq:thetapsi}).

In the following we restrict to the Kerr space-time 
with an observer in the equatorial plane, $\vartheta_O=\tfrac{\pi}{2}$.
Even in this case, a formula for the angular diameters of the shadow
was not known before, as far as we know. In the general case, the angular 
diameters can be calculated analogously; it is true that then the radius 
values $r_{h_i}$ and $r_{v}$ are zeros of a polynomial of higher than 
fourth order, so they cannot be determined in closed form. In terms of 
these radii, however, one gets analytical formulas for the angular 
diameters also in the general case.
The horizontal angular radii $\varrho_{h_i}$ are characterized by 
$\psi _{h_i}=\pm\tfrac{\pi}{2}$, so we must solve the equation 
$1= \sin^2\psi(r_h) = P^2(r_h)$ which in the Kerr case simplifies to 
(use Eq.~\eqref{eq:psi} with Eq.~\eqref{eq:LEKEc})
\begin{gather}
	r_h(r_h-3m)^2 = 4ma^2 \kern4cm \\[\smallskipamount]
	\begin{subequations}
	\label{eq:rh}
	\begin{align}
		\Rightarrow\quad
		r_{h_1} &= 2m + 2m \cos(\zeta/3), \\
		r_{h_2} &= 2m - m \cos(\zeta/3) - \sqrt{3}m \sin(\zeta/3), \\
		r_{h_3} &= 2m - m \cos(\zeta/3) + \sqrt{3}m \sin(\zeta/3),
	\end{align}
	\end{subequations}
\end{gather}
where $\zeta=\arg\bigl((2a^2 m-m^3) - i (2am\sqrt{m^2-a^2})\bigr)$.
Here we have to choose the solutions $r_{h_1}$ and $r_{h_2}$ which
are the radii of the two circular photon orbits in the exterior 
photon region.
Evaluating $P T$ for $r_{h_1}$ and $r_{h_2}$ yields 
by \eqref{eq:AngDia_h} the horizontal angular diameter $\delta_h$ 
of the shadow.

The vertical angular radius corresponds to those boundary points 
where the tangent is horizontal. By \eqref{eq:AngDia_v} we have
$f(r_v) :=\sin^2\varrho_v = \bigl(1-P^2(r_v)\bigr)T^2(r_v)$, so 
the tangent is horizontal if $\frac{\drm f}{\drm r_v}(r_v) = 0$. 
This yields 
\begin{align}
	0 &= (1-P^2) T' - PP'T \, \vert_{r_v} \\
	&= \frac{\sqrt{\Delta(r_O)}\, r_v \, \bigl(r_v (2a^2 + r_O^2) - 3m r_O^2\bigr)
		\bigl(a^2m - r_v(3m^2 - 3m r_v + r_v^2)\bigr)}{a^2\sqrt{\Delta(r_v)} 
			\bigl(r_v(2a^2+r_O^2+r_v^2)-m(r_O^2+3r_v^2)\bigr)^2}
\end{align}
where we have to choose the unique solution inside the 
exterior photon region
\begin{equation}
	r_{v} = \frac{3m r_O^2}{2a^2 + r_O^2}.
	\label{eq:rv}
\end{equation}
With this value $r_v$ we get an analytic expression of the vertical 
angular radius 
\begin{align}
	\sin^2\varrho_v = (1-P^2) & T^2 \vert_{r_v}
		= \frac{27m^2 r_O^2 \bigl(a^2+ r_O (r_O-2 m)\bigr)}{
		r_O^6 + 6a^2 r_O^4+ 3a^2 (4a^2 - 9m^2) r_O^2 + 8a^6 }.
	\label{eq:AngRad_v}
\end{align}
For $a=0$, we recover from (\ref{eq:AngRad_v}) Synge's formula 
\cite{Synge.1966} for a Schwarzschild black hole,
\begin{equation}
	\sin^2 \varrho = \frac{27 m^2 (r_O - 2m)}{r_O^{3}}. 
\label{eq:Synge}
\end{equation}
Since the shadow of a non-rotating black hole is always circular,
the horizontal angular radii $\varrho_{h_i}$ are also given by
\eqref{eq:Synge} in this case.\footnote{For 
$a=0$ one finds $\zeta=\arg(-m^3)=-\pi$ and $r_{h_{1,2}}=3m$. 
Then $T^2(3m)$ reproduces \eqref{eq:Synge}.}
Note that for all values $0 \le a^2 \le m^2$ (\ref{eq:AngRad_v})
gives the same value as (\ref{eq:Synge}), $27 m^2 /r_O^2$, if $m$ is 
negligibly small in comparison to $r_O$. This means that for observers 
far away from the black hole the vertical diameter of the shadow is 
independent of $a$.

In the extremal Kerr space-time, $a=m$, the circular photon orbits are at
$r_{h_1}=4m$ and $r_{h_2}=m$ since $\zeta=\arg(m^3)=0$. Together with 
\eqref{eq:rv} this results in the following formulas for the angular radii
\begin{gather}
\begin{aligned}
	\sin^2 \varrho_{h_1} &= \frac{64 m^2 (r_O - m)^2}{(r_O^2 + 8m^2)^2}, \\
	\sin^2 \varrho_{h_2} &= \frac{m^2 }{(r_O + m)^2},
\end{aligned}
\qquad
	\sin^2 \varrho_{v} = \frac{27m^2 r_O^2}{(r_O + m)^2 (r_O^2 + 8m^2)}.
\label{eq:AngRad_am}
\end{gather}

Finally, we use \eqref{eq:Synge} and \eqref{eq:AngRad_am} to determine the 
angular diameters given by \eqref{eq:AngDia_h} and \eqref{eq:AngDia_v}
for the shadow of the black hole in the center of our Galaxy near Sgr~A* 
and of that in M87. The resulting values are given in Table~\ref{tab:angdia}
together with the corresponding values for the mass $M$ (in multiples of 
the Solar mass $M_{\odot}$) and the distance $r_O$ of the black holes. We 
use two sets of parameters for M87 because the mass estimation based on the 
modeling of stellar dynamics yields a mass twice as big as the estimation 
based on gas dynamical measurements, compare 
Refs.~\citen{Broderick.2015,KormendyHo.2013,GebhardtAdams.2011,WalshBarth.2013}.

The horizontal angular diameter for the extremal rotating black holes is
always about $13\%$ smaller than for the Schwarzschild case while the 
vertical angular diameters $\delta_v$ coincide in all cases. We have 
already observed that the latter is a consequence of the fact that $r_O$
is large in comparison to $m$. It turns out that the shadow of the 
black hole in M87 is not much smaller than that of the black hole 
at the center of our Galaxy; the bigger distance of M87 is almost
compensated by its bigger mass.

\begin{table}[htbp]
\tbl{Horizontal and vertical angular diameter $\delta_h$, $\delta_v$ of the 
shadow for Sgr~A* and M87 for a non-rotating Schwarzschild model 
($a=0$) or an extremal rotating Kerr model ($a=m$) of their black holes\hfill\ }%
{	\begin{tabular}{c|cc|cc|cc} 
		& \multicolumn{2}{c|}{Sgr A*} & 
			\multicolumn{2}{c|}{M87} & \multicolumn{2}{c}{M87} \\
		& $\delta_h$ & $\delta_v$ & 
			$\delta_h$ & $\delta_v$ & $\delta_h$ & $\delta_v$  \\
	\colrule
		$a=0$ & $53.1\unit{\text{\textmu}as}$ & $53.1\unit{\text{\textmu}as}$ &
			$37.8\unit{\text{\textmu}as}$ & $37.8\unit{\text{\textmu}as}$ &
			$20.1\unit{\text{\textmu}as}$ & $20.1\unit{\text{\textmu}as}$ \\
		$a=m$ & $46.0\unit{\text{\textmu}as}$ & $53.1\unit{\text{\textmu}as}$ &
			$32.8\unit{\text{\textmu}as}$ & $37.8\unit{\text{\textmu}as}$ &
			$17.4\unit{\text{\textmu}as}$ & $20.1\unit{\text{\textmu}as}$ \\
	\colrule
		& \multicolumn{2}{l|}{$M = 4.31 \times 10^6 M_{\odot}$,} &
			\multicolumn{2}{l|}{$M = 6.2 \times 10^9 M_{\odot}$,} &
			\multicolumn{2}{l}{$M = 3.5 \times 10^9 M_{\odot}$,} \\
		$m = \frac{M G}{c^2}$ & \multicolumn{2}{l|}{$r_O = 8.33\unit{kpc}$
		\cite{GhezSalim.2008,GillessenEisenhauerEtAl.2009}} &
			\multicolumn{2}{l|}{$r_O = 16.68\unit{Mpc}$
			\cite{Broderick.2015,KormendyHo.2013,GebhardtAdams.2011}} &
			\multicolumn{2}{l}{$r_O = 17.9\unit{Mpc}$
			\cite{Broderick.2015,WalshBarth.2013}}
	\end{tabular}
\label{tab:angdia}}
\end{table}

\section{Conclusions and Outlook}
We have seen that knowing the photon region surrounding a black hole
is essential for calculating the shadow. For both the photon region
and the boundary curve of the shadow we have found analytical formulas
in the general type D class of Pleba{\'n}ski--Demia{\'n}ski space-times.
Since these space-times are not in general asymptotically flat and 
possess additional cosmological horizons, it is not possible to
restrict to observers at infinity as it was done in many other articles 
on shadows of black holes. We have placed our observer at any 
Boyer--Lindquist coordinates in the outer domain of communication 
instead, and we have calculated the shadow for the case that the
four-velocity of the observer is adapted to the symmetries of the 
space-time in the sense that it lies in the intersection of the 
$t$-$\varphi$-plane with the plane spanned by the principal null
directions. Interestingly, for such an observer the shadow
is always symmetric with respect to a horizontal axis, independently 
of an acceleration $\alpha \neq 0$, an inclination $\vartheta _O
\neq \pi/2$ of the observer, or a gravitomagnetic NUT charge $\ell \neq 0$. 
The boundary curve of the shadow depends on the space-time parameters 
$m$, $a$, $\ell$, $\beta$, $\alpha$, and $\Lambda$, as well as on the 
observer's position $(r_O,\vartheta_O)$. For an observer whose four-velocity
is not adapted to the symmetries of the space-time, the boundary
curve of the shadow also depends on the 3 components of the  spatial
velocity with respect to our standard observer. 

Although the acceleration parameter does not destroy the symmetry of the 
shadow with respect to a horizontal axis, it does have such an effect on 
the photon region, the ergosphere and the causality violating region. 
The photon region is reflected at the equatorial plane if the sign of 
$\alpha$ is changed whereas the ergoregion and the causality violating region 
are reflected at the equatorial plane if the sign of $a\alpha$ is changed.

Our estimates of the angular diameters for the shadows of the black holes
in the centers of our Galaxy and of M87 show that the shadows are roughly 
of the same size. Hence the planned observations may provide us with 
shadow images not only of the black hole in our Galaxy but also of that  
in M87. If the current attempts of observing the shadow are successful,
this will give us a chance to deduce the parameters of the black
hole from the boundary curve of the shadow. Our analytical 
formula combined with a Fourier analysis should be a promising 
tool for achieving this goal. We are planning to investigate this 
in a follow-up article.

\section*{Acknowledgments}
Our thanks for helpful discussions go to Nico Giulini, Norman G\"urlebeck, 
David Kofron, Eva Hackmann, and Eugen Radu.
We gratefully acknowledge support from the DFG within the Research 
Training Group 1620 \blqq Models of Gravity\brqq.

%
%
%



\begin{thebibliography}{00}  
\providecommand \enquote [1]{``#1''}%

%
%
%
%
%
%

\bibitem{GrenzebachPerlick.2014}%
	A. Grenzebach, V. Perlick, and C. L{\"a}mmerzahl, %
	{\it Phys. Rev. D} {\bf 89} (2014) 124004%
%
\bibitem{JamesTunzelmann.2014}%
	O. James, E. v. Tunzelmann, P. Franklin, K. S. Thorne, %
	{\it Class. Quantum Grav.} {\bf 32} (2015) 065001%
%
\bibitem{DoelemanWeintroub.2008}%
	S.~S. Doeleman {\it et al.}, %
	{\it Nature} {\bf 455} (2008) 78%
\bibitem{EckartGenzel.1996}%
	A.~Eckart and R.~Genzel, %
	{\it Nature} {\bf 383} (1996) 415%
\bibitem{GillessenEisenhauerEtAl.2009}%
	S.~Gillessen {\it et al.}, %
	{\it Astrophys. J.} {\bf 692} (2009) 1075%
\bibitem{GhezSalim.2008}%
	A.~M. Ghez \textit{et al.}, %
	{\it Astrophys. J.} {\bf 689} (2008) 1044%
\bibitem{HuangCai.2007}%
	L.~Huang, M.~Cai, Zh.-Q.~Shen, and F.~Yuan, %
	{\it Month. Not. R. Astron. Soc.} {\bf 379} (2007) 833%
\bibitem{FalckeMelia.2000}%
	H.~Falcke, F.~Melia, and E.~Agol, %
	{\it Astrophys. J.} {\bf 528} (2000) L13%
\bibitem{KardashevNovikov.2014}%
	N. S. Kardashev, I. D. Novikov, V. N. Lukash, S. Pilipenko, et. al., %
	{\it Uspekhi Fizicheskih Nauk} {\bf 184} (2014) pp.~1319, %
	english transl. {\it arXiv:1502.06071}%
%
\bibitem{Levi-Civita.1919}%
	T. Levi-Civita, %
	{\it Rend. Accad. Lincei} {\bf 28(1)} (1919) pp.~3%
\bibitem{Weyl.1917}%
	H. Weyl, %
	{\it Ann. Physik} {\bf 54 (359)} (1917) pp.~117%
\bibitem{Weyl.1919}%
	H. Weyl, %
	{\it Ann. Physik} {\bf 59 (364)} (1919) pp.~185%
\bibitem{EhlersKundt.1962}
	J. Ehlers and W. Kundt, %
	Exact Solutions of the Gravitational Field Equations, %
	in {\it Gravitation: an introduction to current research}, %
	ed. L. Witten %
	(John Wiley \&\ Sons, New York, 1962), chp.~2%
\bibitem{HongTeo.2005}
	K. Hong and E. Teo, %
	{\it Class. Quantum Grav.} {\bf 22} (2005) pp.~109%
\bibitem{GriffithsPodolsky.2009}%
	J.~B. Griffiths and J.~Podolsk{\'y}, %
	{\it Exact Space-Times in Einstein's General Relativity}, 
	(Cambridge University Press, Cambridge, 2009)%
\bibitem{HongTeo.2003}
	K. Hong and E. Teo, %
	{\it Class. Quantum Grav.} {\bf 20} (2003) pp.~3269%
\bibitem{KinnersleyWalker.1970}
	W. Kinnersley and M. Walker %
	{\it Phys. Rev. D} {\bf 2} (1970) pp.~1359%
\bibitem{Bonnor.1983}
	W. B. Bonnor, %
	{\it Gen. Rel. Grav.} {\bf 15} (1983) pp.~535%
\bibitem{BonnorDavidson.1992}
	W. B. Bonnor and W. Davidson, %
	{\it Class. Quantum Grav.} {\bf 9} (1992) pp.~2065%
%
\bibitem{Synge.1966}%
	J.~L. Synge, %
	{\it Mon. Not. R. Astron. Soc.} {\bf 131} (1966) 463%
\bibitem{Bardeen.1973}%
	J.~M. Bardeen, %
	in {\it Black Holes (Les Astres Occlus)}, 
	eds. C.~DeWitt and B.~S.~DeWitt (Gordon and Breach, New York, 1973), p.~215%
\bibitem{HiokiMaeda.2009}%
	K.~Hioki and K.-I. Maeda, %
	{\it Phys. Rev. D} {\bf 80} (2009) 024042%
\bibitem{BardeenCunningham.1973}%
	J.~M. Bardeen and C.~T. Cunningham, %
	{\it Astrophys. J.} {\bf 183} (1973) 237%
\bibitem{Luminet.1979}%
	J.-P. Luminet, %
	{\it Astron. Astrophys.} {\bf 75} (1979) 228%
\bibitem{DexterAgol.2012}%
	J.~Dexter, E.~Agol, P.~C. Fragile, and J.~C. McKinney, %
	{\it J. Phys.: Con. Ser.} {\bf 372} (2012) 012023%
\bibitem{YounsiWu.2012}
	Z. Younsi, K. Wu, and S. V. Fuerst, %
	{\it A\& A.} {\bf 545} (2012) A13%
\bibitem{MoscibrodzkaFalcke.2014}%
	M.~Mo{\'s}cibrodzka, H. Falcke, H.~Shiokawa, and C.~F. Gammie, %
	{\it A\& A.} {\bf 570} (2014) A7%
\bibitem{MoscibrodzkaShiokawa.2012}%
	M.~Mo{\'s}cibrodzka, H.~Shiokawa, C.~F. Gammie, and J.~C. Dolence, %
	{\it Astrophys. J. Lett.} {\bf 752} (2012) L1%
\bibitem{DexterFragile.2013}%
	J.~Dexter and 
	P.~C. Fragile, %
	{\it Mon. Not. R. Astron. Soc.} {\bf 432} (2013) 2252%
\bibitem{Debever.1971}%
	R. Debever, %
	{\it Bull. Soc. math Belgique} {\bf 23} (1971) 360%
\bibitem{PlebanskiDemianski.1976}%
	J.~F. Pleba{\'n}ski and M.~Demia{\'n}ski, %
	{\it Ann. Phys.} {\bf 98} (1976) 98%
\bibitem{StephaniKramer.2003}%
	H. Stephani, D.~Kramer, M.~MacCallum, C.~Hoenselaers, E.~Herlt, %
	{\it Exact Solutions of Einstein's Field Equations}, 
	(Cambridge University Press, Cambridge, 2003)%
\bibitem{AlievGumrukcuoglu.2005}%
	A.~N. Aliev and A.~E. G{\"u}mr{\"u}k{\c c}{\"u}o{\u g}lu, %
	{\it Phys. Rev. D} {\bf 71} (2005) 104027%
\bibitem{MankoRuiz.2005}%
	V.~S. Manko and E.~Ruiz, %
	{\it Class. Quantum Grav.} {\bf 22} (2005) 3555%
\bibitem{Plebanski.1975}%
	J.~F. Pleba{\'nski}, %
	{\it Ann. Phys.} {\bf 90} (1975) 196%
\bibitem{Carter.1968b}%
	B.~Carter, %
	{\it Commun. Math. Phys.} {\bf 10} (1968) 280%
\bibitem{Perlick.2004}%
	V.~Perlick, %
	{\it Living Rev. Relativ.} {\bf 7} (2004) 9%
\bibitem{ONeill.1995}%
	B.~O'Neill, %
	{\it The Geometry of Kerr Black Holes} 
	(A~K~Peters, Wellesley, 1995)%
\bibitem{Grenzebach.2015}%
	A.~Grenzebach, %
	Aberrational Effects for Shadows of Black Holes, %
	to appear in {\it Proc. 524th WE-Heraeus-Seminar %
	\blqq Equations of Motion in Relativistic Gravity\brqq}, %
	eds. D. Puetzfeld, C. L\"ammerzahl, B.F. Schutz %
	(Springer, Heidelberg, 2015)
\bibitem{Broderick.2015}%
	A.~Broderick, R.~Narayan, J.~Kormendy, {\it et al.}, %
	The Event Horizon of M87, %
	submitted to {\it Astrophys. J.}
\bibitem{KormendyHo.2013}%
	J.~Kormendy and L.~C.~Ho, %
	{\it Annu. Rev. Astron. Astrophys.} {\bf 51} (2013) 511%
\bibitem{GebhardtAdams.2011}%
	K.~Gebhardt, J.~Adams, D.~Richstone, {\it et al.}, %
	{\it Astrophys. J.} {\bf 729} (2011) 119%
\bibitem{WalshBarth.2013}%
	J.~L.~Walsh, A.~J.~Barth, L.~C.~Ho, and M.~Sarzi, %
	{\it Astrophys. J.} {\bf 770} (2013) 86%
\end{thebibliography}
\end{document}